\documentclass[
    aps,
    prl,
    reprint,
    superscriptaddress,
    nobibnotes,
    preprintnumbers
]{revtex4-2}

\usepackage{amsmath,amssymb}
\usepackage{bm}
\usepackage{graphicx}
\usepackage{dcolumn}
\usepackage{booktabs}
\usepackage{xcolor}
\usepackage{hyperref}
\usepackage{physics}
\usepackage{orcidlink}
\usepackage{tikz}
\usetikzlibrary{arrows.meta,decorations.markings}

\hypersetup{
    colorlinks=true,
    linkcolor=blue!60!black,
    citecolor=blue!60!black,
    urlcolor=blue!60!black
}

\usepackage{placeins}

\renewcommand{\dd}{\mathrm{d}}
\newcommand{\ii}{\mathrm{i}}

\begin{document}

\preprint{IPPP/26/80}

\title{Fermion Scattering on Quantum Computers}

\author{Steven Abel\,\orcidlink{0000-0003-1213-907X}}
\affiliation{Institute for Particle Physics Phenomenology, Durham University, Durham DH1 3LE, UK\\[0.35em]}

\author{Michael Spannowsky\,\orcidlink{0000-0002-8362-0576}}
\affiliation{Institute for Theoretical Physics, Campus S\"ud, Karlsruhe Institute of Technology, D-76128 Karlsruhe, Germany\\[0.35em]}
\affiliation{\mbox{Institute for Quantum Materials and Technologies, Karlsruhe Institute of Technology, Karlsruhe 76131, Germany}}

\author{Simon Williams\,\orcidlink{0000-0001-8540-0780}}
\email[Corresponding author: ]{simon.williams@kit.edu}
\affiliation{Institute for Theoretical Physics, Campus S\"ud, Karlsruhe Institute of Technology, D-76128 Karlsruhe, Germany\\[0.35em]}
\affiliation{\mbox{Institute for Quantum Materials and Technologies, Karlsruhe Institute of Technology, Karlsruhe 76131, Germany}}

\begin{abstract}
Resonant scattering links particle interactions to the formation and decay of unstable excitations. We develop a hybrid quantum-computing framework to study these dynamics through fermion--antifermion collisions. Our formulation of $(1+1)$-dimensional Yukawa theory encodes fermions in qubits and the scalar field in qumodes. We identify the operations needed for a photonic implementation and benchmark the dynamics with tensor-network simulations. Near resonance, we resolve the build-up of a localised scalar excitation and the subsequent partial return of energy to the fermions. Enhanced entanglement accompanies this exchange. Time-dependent correlator measurements reproduce the analytically predicted decay rate to a few percent precision, with  energy scans exhibiting a resonance peak consistent with packet-averaged Breit--Wigner behaviour. These results establish a concrete scattering benchmark for hybrid quantum computers and demonstrate how local real-time observables can probe particle-resonance dynamics.
\end{abstract}

\maketitle

\section{Introduction}

Real-time scattering provides a demanding test of our ability to simulate interacting quantum fields. Quantum algorithms offer access to dynamics that are difficult to reconstruct with Euclidean lattice methods, while tensor networks provide classical reference calculations for developing and assessing quantum simulations. Qubit-based~\cite{Jordan:2012QFT} and continuous-variable~\cite{Marshall:2015CVQFT} approaches have been developed for quantum field theory (QFT), alongside studies of real-time scattering~\cite{Abel:2025zxb,Ingoldby:2025bdb} and vacuum decay~\cite{Abel:2025pxa, Pavesic:2025nwm}. Treating bosons and fermions together is particularly important: their distinct statistics favour different local representations, while their interactions generate energy exchange and entanglement.

Tensor-network calculations have resolved spectra in Abelian and non-Abelian lattice gauge theories~\cite{Banuls:2013SchwingerSpectrum,Banuls:2017SU2Basis}, connected gauge-field representations to atomic quantum simulation~\cite{Rico2014TensorNetworks}, and accessed real-time thermal correlators~\cite{Banuls:2020RealTimeThermalQFT}. They also describe string breaking~\cite{Pichler2016RealTime} and elastic and inelastic meson scattering~\cite{Belyansky:2024Collisions}. Recent work has investigated time-domain resonance observables in glueball scattering and demonstrated a related toy collision model on quantum hardware~\cite{Balaji:2026TimeDomain}.

Yukawa theory supplies a complementary setting with explicit fermionic matter coupled to a dynamical scalar field. It allows us to follow the scalar response to a fermion--antifermion collision and the accompanying redistribution of energy and entanglement. Previous quantum simulations include digital studies of a single-site Yukawa model~\cite{Kaldenbach:2024Yukawa}, spin--phonon experiments on small spatial lattices~\cite{Than:2025SpinPhonon}, and trapped-ion simulations of the static model~\cite{Wang:2026StaticYukawa}.

Our contribution is a spatially resolved study of resonant fermion--antifermion collisions that connects a hybrid qubit--oscillator encoding of lattice Yukawa theory to matrix-product-state (MPS) reference calculations. We track local scalar excitation, energy exchange between particle sectors and entanglement growth, and examine how finite wave packets and observation windows shape the resonance response. This combination provides a scattering test case for hybrid quantum simulation and identifies the distinction between finite-time fitted widths and an intrinsic decay rate. The encoding also specifies the operations needed for a photonic implementation. In $(1+1)$ dimensions, the continuum Hamiltonian is
 \begin{align}
H ~= ~ \int dx\,\Bigg\{&
  \frac{1}{2}{\pi}^{\,2}
  +\frac{1}{2}\bigl(\partial_x{\phi}\bigr)^2
  +\frac{1}{2}m_\phi^2{\phi}^{\,2}
 \label{eq:yukawa_hamiltonian_continuum}
 \\
+\,\bar \psi &
  \left(-i\gamma_1\,\partial_x+m_f\right)\psi
  +g{\phi}
  \left(\bar{\psi}\psi-\mathcal{C}(x)\right)
\Bigg\}~,\nonumber
\end{align}
where  we set $c=\hbar=1$, where it is convenient to take $\gamma_0 = \sigma_z$ and $\gamma_1 = i \sigma_y$, and where the constant $\mathcal{C}$ may be added to implement normal ordering, and where $ \pi = \partial_0 \phi$.  One could trivially add more general scalar potentials, for example additional  $\lambda \phi^4$ terms, but in this work we will not consider them.

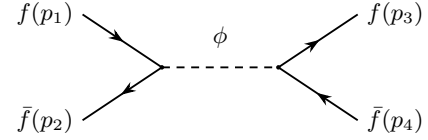
\begin{figure}[tb]
\centering
\begin{tikzpicture}[
  scale=0.7,
  line width=0.7pt,
  fermion/.style={
    decoration={markings,
      mark=at position 0.52 with {
        \arrow{Stealth[length=2mm,width=1.4mm]}
      }},
    postaction={decorate}
  },
  scalar/.style={dash pattern=on 3pt off 2.5pt}
]
  \coordinate (L) at (0,0);
  \coordinate (R) at (2.2,0);

  \draw[fermion]
    (-1.5,1.0) node[left] {$f(p_1)$} -- (L);
  \draw[fermion]
    (L) -- (-1.5,-1.0) node[left] {$\bar f(p_2)$};

  \draw[scalar]
    (L) -- node[above=4pt] {$\phi$} (R);

  \draw[fermion]
    (R) -- (3.7,1.0) node[right] {$f(p_3)$};
  \draw[fermion]
    (3.7,-1.0) node[right] {$\bar f(p_4)$} -- (R);

  \fill (L) circle (1.2pt);
  \fill (R) circle (1.2pt);
\end{tikzpicture}
\caption{Tree-level scalar-mediated $s$-channel contribution
to $f\bar f$ scattering. Arrows indicate fermion-number flow.}
\label{fig:ffbar-phi-tree}
\end{figure}

\begin{figure}[h]
    \centering
    \includegraphics[width=\columnwidth]{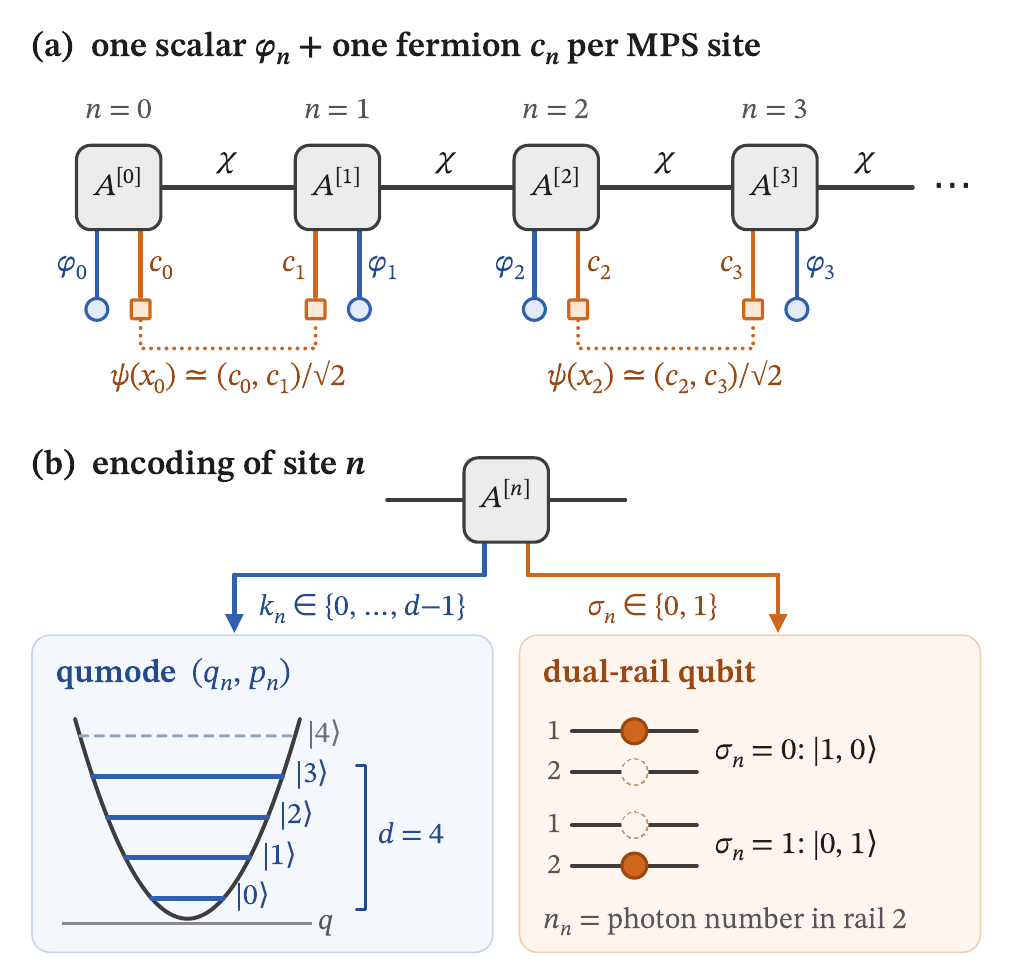}
    \caption{Lattice Yukawa model and photonic encoding. (a) Schematic of the one-dimensional lattice representation, with each MPS site containing one scalar degree of freedom, $\phi_n$, and one staggered fermionic mode, $c_n$. The tensors $A^{[n]}$ are connected by bonds of dimension $\chi$, whilst pairs of neighbouring staggered fermionic modes reconstruct the two components of the continuum fermion field, $\psi(x_0)\simeq(c_0,c_1)/\sqrt{2}$. (b) Encoding of the local MPS degree of freedom on the photonic device. The physical index of $A^{[n]}$ factorises into a truncated scalar occupation $k_n\in\{0,\ldots,d-1\}$, represented by a qumode with quadratures $(q_n,p_n)$, and a fermionic occupation $\sigma_n\in\{0,1\}$, represented by a dual-rail photonic qubit. The qumode truncation $d=4$ is shown for illustration.}
    \label{fig:schematic}
\end{figure}

Near the scalar resonance, fermion--antifermion scattering is dominated by the $s$-channel contribution shown in
Fig.~\ref{fig:ffbar-phi-tree}, with the vertices contributing factors of $-i g$. 
Provided the scalar is massive enough, it will produce a characteristic `resonance peak' in the scattering cross-section that we can use to probe its properties, proportional to the Breit-Wigner (BW) form. This is given by 
\begin{equation}
|\mathcal M_s|^2 ~\simeq~
\frac{g^4(E^2-4M_f^2)^2}
{(E^2-M_\phi^2)^2+M_\phi^2\Gamma^2}~,
\label{eq:yukawa_bw_amplitude_squared}
\end{equation}
where $M_f$ and $M_\phi$ are the pole masses of the fermion and scalar. 
Here $E$ is the centre-of-mass energy, and we use tree-level masses as a leading-order benchmark. The leading scalar decay rate is given by the $\phi\to f\bar f$ contribution,
\begin{equation}
\Gamma_{\rm cont}
~=~
\frac{g^2}{2m_\phi}\, 
\sqrt{1-\frac{4m_f^2}{m_\phi^2}}\,
\Theta(m_\phi-2m_f)~.
\label{eq:yukawa_width_continuum}
\end{equation}
In this letter we will focus on reproducing this resonance, and will  demonstrate that the expected result can be recovered to remarkable precision (roughly 2\%). 


\section{Encodings and methodology}

\begin{figure*}[t]
    \centering
    \includegraphics[width=\textwidth]{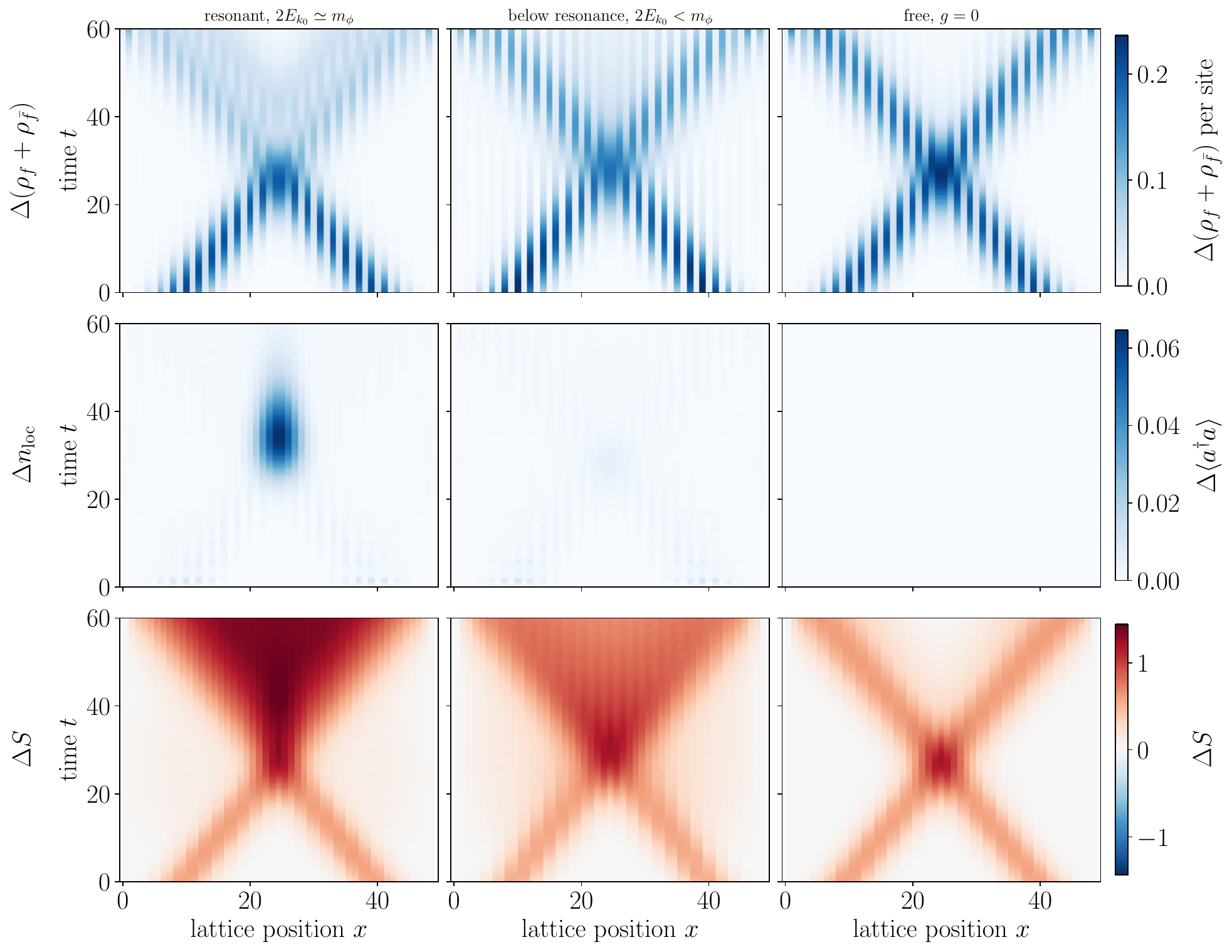}
    \caption{Real-time fermion–scalar scattering in the lattice Yukawa model. Space–time evolution for an interacting resonant collision, $2E_{k_0}\simeq m_\phi$ (left), an interacting collision below resonance, $2E_{k_0}<m_\phi$ (centre), and the free theory, $g=0$ (right), for $N=50$ lattice sites. Top: vacuum-subtracted fermion and antifermion density, $\Delta(\rho_f+\rho_{\bar f})$. Middle: change in the local scalar occupation, $\Delta n_{\rm loc}=\Delta\langle a^\dagger a\rangle$. Bottom: change in the bipartite entanglement entropy, $\Delta S$. On resonance, the collision produces a pronounced, spatially localised scalar excitation that persists after the incoming fermion wave packets overlap, accompanied by enhanced entanglement growth. The scalar response is strongly suppressed below resonance and absent in the free theory.}
    \label{fig:scattering}
\end{figure*}

The two kinds of particles, scalars and fermions, appearing in the Yukawa model naturally require different quantum encodings. In $(1+1)$ dimensions, the fermion field is a two-component Dirac spinor, $\psi=(\psi_1,\psi_2)^T$. Here, we use a staggered discretisation in which these components are carried by neighbouring lattice sites,
\begin{equation}
\psi(x_{2j})\simeq\frac{1}{\sqrt{2a}}
\begin{pmatrix}
c_{2j}\\ c_{2j+1}
\end{pmatrix}~,
\end{equation}
where $c_n$ obey the fermionic anticommutation relations~\cite{KogutSusskind1975,Susskind1977}, and $a$ is the lattice spacing. Henceforth we set $a=1$. Each staggered mode is therefore either empty or occupied, thus requiring a two-level register. By contrast, the scalar field has a continuous field amplitude and conjugate momentum at each site, which can be represented directly by the quadrature operator of the Hilbert space of a local oscillator, known as a qumode~\cite{Abel:2024kuv}.

This field structure naturally motivates a hybrid framework of continuous and discrete variable degrees of freedom, which we will refer to as a CV-DV architecture. Related oscillator--qubit frameworks have been developed for superconducting and trapped-ion platforms~\cite{Crane:2026Hybrid,Araz:2024hybrid}. Here, we identify photonic quantum computers as a natural platform, as they allow for qumodes and qubits to be encoded on the same quantum device. In the CV computational paradigm of photonic quantum computers, each scalar degree of freedom is encoded in a qumode and each fermionic occupation as a dual-rail photonic qubit, as shown in Fig.~\ref{fig:schematic}. Fermionic statistics are imposed through a Jordan-Wigner (JW) transformation. In (1+1) dimensions, the associated parity strings cancel from the nearest-neighbour hopping terms on the lattice with open boundary conditions (OBC). However, the non-local strings remain in spatially separated fermionic operators and wavepacket observables. For classical emulation, or a fully digital qubit implementation, the local scalar Hilbert space can instead be truncated to $d$ states.

Latticising the Hamiltonian from Eq.~\eqref{eq:yukawa_hamiltonian_continuum} over $N$ sites yields, 
\begin{align}
H~=~{}&\sum_{n=0}^{N-1}
\left[
\frac{\hat p_n^2}{2}
+\frac{m_\phi^2+z_n}{2}\hat q_n^2
\right]
\,-\,\sum_{n=0}^{N-2}\hat q_n\hat q_{n+1}
\nonumber\\
&-\,\frac{\ii}{2}\sum_{n=0}^{N-2}
\left(c_n^\dagger c_{n+1}-c_{n+1}^\dagger c_n\right)
\,+\, m_f\sum_{n=0}^{N-1}(-1)^n c_n^\dagger c_n
\nonumber\\
&\,+\,g\sum_{n=0}^{N-1}\hat q_n
\left[(-1)^n c_n^\dagger c_n-C_n\right]~,
\label{eq:yukawa_hamiltonian}
\end{align}
where we make the identification of Ref.~\cite{Abel:2025zxb}, $\hat q_n=\,\phi(x_n)$ and $\hat p_n=
\pi(x_n)$. Here $z_n$ counts the neighbouring sites, taking the value $1$ at the lattice boundary, and $2$ in the interior. The subtraction of the term $C_n=(-1)^n(P_-)_{nn}$ normal-orders the fermion bilinear with respect to the free vacuum, where $P_\pm$ projects onto the positive/negative-frequency modes of the free lattice Dirac Hamiltonian. 

The pure scalar terms generate entirely Gaussian dynamics which can easily be realised on continuous variable devices. Meanwhile fermionic hopping simply requires entangling gates between neighbouring qubits. By contrast the Yukawa term generates a displacement of the scalar momentum controlled by the fermionic occupation. This requires an onsite non-Gaussian operation in the photonic implementation; measurement-based methods for generating non-Gaussian operations are discussed in Refs.~\cite{Sabapathy:2019ybd, Abel:2024kuv}. 

We calculate the dynamics using matrix-product-state (MPS) simulations~\cite{Schollwock:2011MPS} of fermion-antifermion scattering on a finite lattice with OBC. The interacting vacuum $\vert \Omega \rangle$ is obtained through imaginary-time evolution using time-evolving block decimation (TEBD)~\cite{Vidal:2004TEBD}, following the procedure of Ref.~\cite{Abel:2025pxa}. We then prepare
\begin{equation}
\vert\Psi_0\rangle\propto b_\alpha^\dagger d_\beta^\dagger\vert\Omega\rangle,
\qquad
b_\alpha^\dagger=\sum_n\alpha_nc_n^\dagger,
\qquad
d_\beta^\dagger=\sum_n\beta_n^*c_n,
\end{equation}
where $b_\alpha^\dagger$ and $d_\beta^\dagger$ create fermion and antifermion wave packets in the free-particle basis, with amplitudes $\alpha_n$ and $\beta^*_n$.

To construct their profiles, we use the free lattice Dirac spinors $u(k)$ and $v(k)$. These satisfy
\begin{equation}
h(k)u(k)=E_ku(k),\qquad
h(-k)v(k)=-E_kv(k),
\end{equation}
with $h(k)=m_f\sigma_z+\sin k\,\sigma_x$ and $E_k=\sqrt{m_f^2+\sin^2k}$. As in the continuum Dirac-field expansion, $u$ multiplies a fermion annihilation operator and $v$ an antifermion creation operator. We form Gaussian trial profiles
\begin{align}
(f_+)_n&=
e^{-(n-x_+)^2/(4\sigma^2)}
e^{\ii k_0n}u_{s(n)}(k_0),
\nonumber\\
(f_-)_n&=
e^{-(n-x_-)^2/(4\sigma^2)}
e^{\ii k_0n}v_{s(n)}(-k_0),
\end{align}
where $s(n)=1$ on even sites and $2$ on odd sites. The normalised profiles are
\begin{equation}
\alpha=\frac{P_+f_+}{\|P_+f_+\|},
\qquad
\beta=\frac{P_-f_-}{\|P_-f_-\|},
\end{equation}
where $P_\pm$ select the positive- and negative-frequency modes of the free open chain. The resulting fermion and antifermion packets have momenta centred on $+k_0$ and $-k_0$, respectively, and move towards one another.

For the interacting simulations we take $m_f=0.5$, $m_\phi=1.732$ and $g=0.6$, retaining $d=4$ local scalar Fock states. Real-time evolution uses second-order TEBD with $\Delta t=0.05$ and maximum bond dimension $\chi=96$, whilst the vacuum is prepared with $\chi=48$.

We resolve the fermion and antifermion populations through
\begin{equation}
\rho_f(n,t)=(P_+GP_+)_{nn},\qquad
\rho_{\bar f}(n,t)=[P_-(I-G)P_-]_{nn},
\end{equation}
where $G_{ij}(t)=\langle c_j^\dagger c_i\rangle_t$. The local scalar occupation is
\begin{equation}
n_{\rm loc}(n,t)=
\frac{\langle\hat p_n^2\rangle_t+
\omega_{\rm loc}^2\langle\hat q_n^2\rangle_t}
{2\omega_{\rm loc}}-\frac12,
\qquad
\omega_{\rm loc}=\sqrt{m_\phi^2+2}.
\end{equation}
We also calculate the von Neumann entropy $S_n$ across each spatial bond, including both scalar and fermionic degrees of freedom. All plotted changes $\Delta$ are measured relative to the corresponding vacuum.


\section{Scattering Experiments}

We perform numerical scattering experiments to resolve the transfer of excitation between the fermionic and scalar sectors. Figure~\ref{fig:scattering} shows the collision of a fermion and an antifermion on an $N=50$ site lattice, with packet width $\sigma=3$ and initial centres $x_+=9.5$ and $x_-=39.5$. We compare an interacting collision at $k_0=0.785$, a lower-energy collision at $k_0=0.35$, and free evolution at $g=0$ with $k_0=0.785$. In the first case, the free carrier energy $2E_{k_0}\simeq1.732$ matches the bare scalar mass, defining the nominal resonant setting. The other two preparations test the energy dependence of the response and its origin in the Yukawa interaction.

As the near-resonant packets overlap, a pronounced scalar excitation develops around the collision centre and persists as the outgoing fermionic density separates. The corresponding scalar response is strongly suppressed in the lower-energy collision and absent in the free evolution. A weaker response is already present before the collision in the interacting runs. This is expected because the packets are constructed from free-particle modes acting on the interacting vacuum, and their subsequent evolution includes dressing and preparation transients. Vacuum subtraction does not remove these contributions. The local oscillator occupation therefore measures excitation of the scalar sector, rather than an asymptotic scalar-particle yield.

The bipartite entropy in the bottom row of Fig.~\ref{fig:scattering} provides a complementary diagnostic. Related diagnostics have been used to characterise entanglement generation in meson scattering~\cite{Rigobello2021Entanglement} and energy loss and entropy production in scattering with matter~\cite{BarataRico2026Jet}. Even in the free theory, the delocalised packets generate entanglement across spatial cuts as they propagate and overlap. With the interaction switched on, the entropy increase is larger and extends through the outgoing state. The strongest enhancement occurs in the nominal resonant preparation, where it accompanies the spatially localised scalar response. These comparisons establish an interaction-induced, energy-selective modification of the collision dynamics.

To quantify the energy transfer accompanying this scalar response, we decompose Eq.~\eqref{eq:yukawa_hamiltonian} as $H=H_f+H_\phi+H_{\rm int}$, where $H_f$ contains the fermion hopping and mass terms, $H_\phi$ the quadratic scalar terms, and $H_{\rm int}$ the Yukawa coupling including its subtraction. The vacuum-subtracted sector energies are
\begin{equation}
\Delta E_X(t)=\langle H_X\rangle_t-\langle H_X\rangle_\Omega,
\qquad X\in\{f,\phi,{\rm int}\}.
\label{eq:sector_energies}
\end{equation}
Their sum gives the conserved excess energy of the prepared state.

\begin{figure}[t]
    \centering
    \includegraphics[width=\columnwidth]{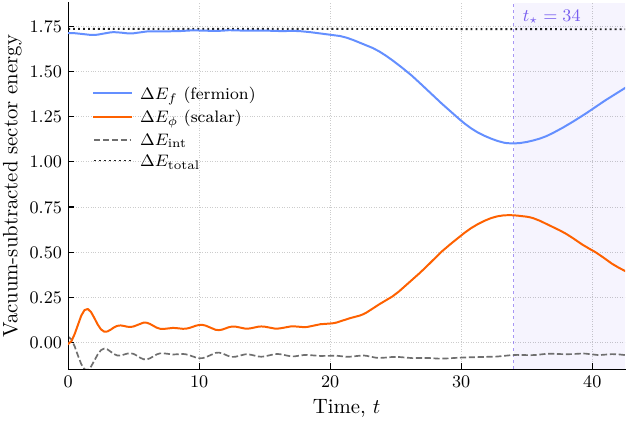}
    \caption{Energy exchange during a near-resonant collision on an $N=50$ site lattice. Vacuum-subtracted fermionic, scalar and interaction energies, together with their sum. The scalar energy reaches its maximum near $t_\star\simeq34$, where the fermionic contribution is smallest. The shaded interval shows the subsequent decrease of the scalar excitation and partial recovery of the fermionic energy. The total energy remains approximately constant, while the interaction contribution is smaller than the energy exchanged between the two sectors.}
    \label{fig:sector_energy}
\end{figure}

Figure~\ref{fig:sector_energy} shows a decrease in the fermionic energy as the scalar energy rises during the collision. The scalar contribution reaches its maximum near $t_\star\simeq34$, after which the energy flow reverses: the scalar excitation decreases while the fermionic contribution recovers. The interaction energy remains smaller than the energy exchanged between the two sectors, and the total energy is approximately constant over the displayed interval. The recovery is partial within this interval. Together with the spatially resolved densities, this behaviour is consistent with the formation of an intermediate scalar excitation followed by energy transfer back into the fermionic sector.

\begin{figure}[t]
    \centering
    \includegraphics[width=\columnwidth]{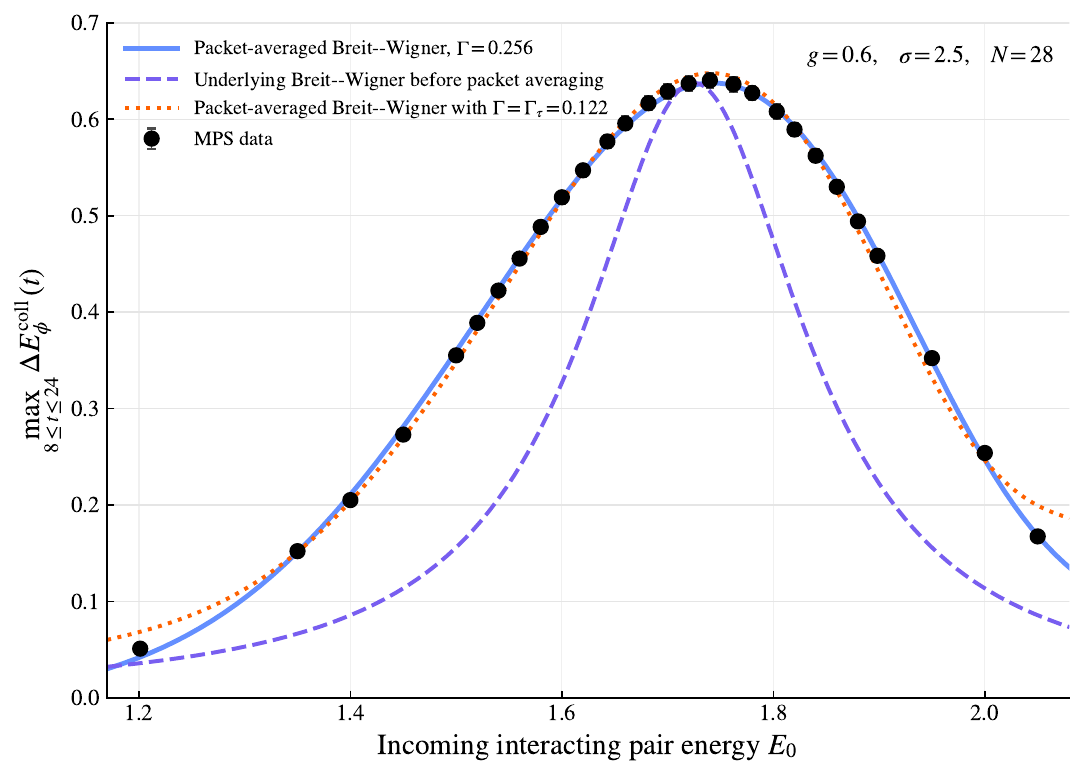}
    \caption{Energy dependence of the scalar response. Maximum collision-induced scalar-energy response over $8\leq t\leq24$ for $g=0.6$, $\sigma=2.5$ and $N=28$. Black points show the MPS results. The solid blue curve is a packet-averaged Breit--Wigner model with fitted width parameter $\Gamma=0.256$; the dashed curve shows the corresponding profile before packet averaging. The dotted orange curve fixes the width to the lifetime reference $\Gamma_\tau=0.122$ before packet averaging. The similar descriptions of the peak illustrate the limited sensitivity of this finite-time observable to the underlying width.}
    \label{fig:bw}
\end{figure}

Having resolved this exchange in time, we probe its energy dependence by scanning the incoming momentum on an $N=28$ site lattice with $\sigma=2.5$ and $g=0.6$. Denoting the incoming interacting pair energy by $E_0$ and the collision-induced scalar-energy response by $\Delta E_\phi^{\rm coll}(t;E_0)$, we consider its maximum during a fixed observation window,
\begin{equation}
R_{\max}(E_0)=\max_{8\leq t\leq24}\Delta E_\phi^{\rm coll}(t;E_0).
\label{eq:maximum_scalar_response}
\end{equation}
The resulting scan, shown in Fig.~\ref{fig:bw}, exhibits a peak near $E_0\simeq1.73$. This observable characterises the finite-time scalar response; it is not itself an asymptotic scattering cross section.

For the staggered lattice that we are working with, 
the scattering is still governed by the Breit-Wigner form of Eq.~\eqref{eq:yukawa_bw_amplitude_squared}
but with dispersion relation appropriate for a lattice density of final states:
$E_k=\sqrt{m_f^2+\sin^2k}$. The leading-order, infinite-volume decay width of a scalar at rest is then 
\begin{equation}
\Gamma_{\rm lat}
~=~\frac{\Gamma_{\rm cont}}{|\cos k_\star|}\,,
\qquad
\sin^2k_\star~=~\frac{m_\phi^2}{4}-m_f^2\,,
\label{eq:yukawa_width_lattice}
\end{equation}
valid within the fermion-pair band, $2m_f<m_\phi<2\sqrt{m_f^2+1}$. (The additional factor arises from the group-velocity Jacobian in the two-body phase space.) For the parameters used, the continuum prediction is $\Gamma_{\rm cont}\simeq0.08485$, whereas the lattice prediction is $\Gamma_{\rm lat}\simeq0.1200$. 

For comparison, we fit a Breit--Wigner profile,
\begin{equation}
R_0(E)~=~A\,\frac{(\Gamma/2)^2}{(E-E_\star)^2+(\Gamma/2)^2}+B(E)\,,
\label{eq:breitwigner}
\end{equation}
where $E_\star$ is the resonance energy, $\Gamma$ is a width parameter, $A$ sets the response scale and $B(E)$ allows for a smooth background. The solid blue line of Fig.~\ref{fig:bw} compares the data with the packet-averaged model.

Allowing the width parameter to vary gives $\Gamma=0.256$ for this maximum-response fit. 
However, the finite momentum distribution of the incoming packets is expected to have broadened the response relative to an underlying `true' BW profile. To determine this true BW profile, we may determine an independent lifetime reference from the normalised connected scalar-source correlator,
\begin{equation}
C_\phi(t)~=~ \frac{ \langle\Omega|\delta Q_0 e^{-i(H-E_\Omega)t} \delta Q_0|\Omega\rangle }{ \langle\Omega|(\delta Q_0)^2|\Omega\rangle },
\label{eq:scalar_survival}
\end{equation}
where $E_\Omega$ is the vacuum energy and
\begin{equation}
\delta Q_0~=~ \frac{1}{\sqrt{N}} \sum_{n=0}^{N-1} \left(\hat q_n-\langle\hat q_n\rangle_\Omega\right).
\end{equation}
After the initial preparation transient and before finite-volume recurrences, we fit this measured complex correlator to
\begin{equation}
C_\phi(t)~\simeq~ Z e^{-iM_\tau t-\Gamma_\tau t/2} +D e^{-i\omega_b t}.
\label{eq:scalar_lifetime_fit}
\end{equation}
Here $Z$ and $D$ are complex amplitudes, and the second term models an oscillatory contribution near the lattice band edge. The survival probability of the pole component decays as $e^{-\Gamma_\tau t}$, giving the lifetime $\tau=1/\Gamma_\tau$.

This analysis gives $\Gamma_\tau\simeq0.122$, numerically close to the leading-order lattice prediction $\Gamma_{\rm lat}\simeq0.1200$. Fixing the underlying Breit--Wigner width to $\Gamma_\tau$ and refitting the remaining parameters gives the orange packet-averaged curve in Fig.~\ref{fig:bw}. It reproduces the broad peak remarkably well. 

The limited width sensitivity of the short-window maximum motivates a complementary analysis of the time-integrated response,
\begin{equation}
I(E_0,t)=\int_0^t\Delta E_\phi^{\rm coll}(t';E_0)\,\dd t'.
\label{eq:integrated_scalar_response}
\end{equation}
At each read-out time, its energy dependence is fitted with a resonance profile and a linear background. This defines an effective width $\Gamma_{\rm fit}(t)$ for the integrated observable, distinct from the width obtained from $R_{\max}$.

\begin{figure}[t]
    \centering
    \includegraphics[width=\columnwidth]{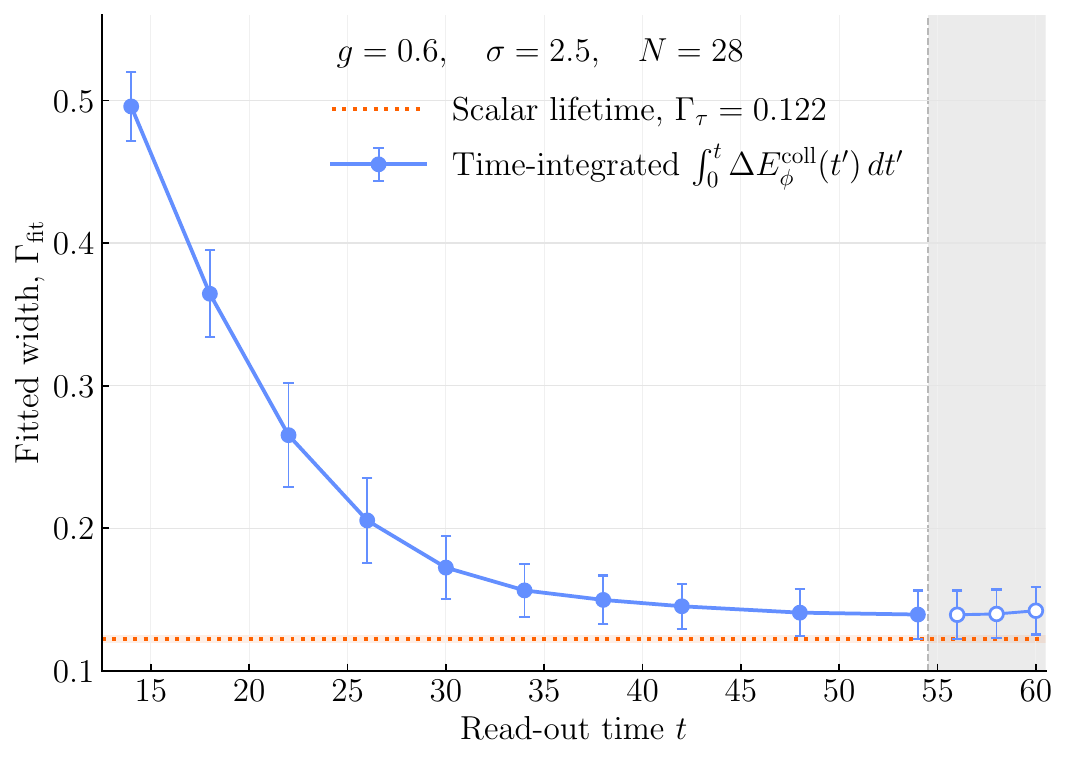}
    \caption{Read-out-time dependence of the fitted width. Width $\Gamma_{\rm fit}$ extracted from the energy dependence of the time-integrated scalar response in Eq.~\eqref{eq:integrated_scalar_response}, for $g=0.6$, $\sigma=2.5$ and $N=28$. The central values decrease towards approximately $0.14$ before the region marked as affected by finite-volume recurrence (grey); open points are excluded from the interpretation. The orange dotted line denotes the lifetime reference $\Gamma_\tau=0.122$. Error bars combine fit-covariance and leave-one-out sensitivity for fits with a linear background.}
    \label{fig:bw_time_dependence}
\end{figure}

Figure~\ref{fig:bw_time_dependence} shows a substantial reduction of $\Gamma_{\rm fit}$ with increasing observation time. The decrease becomes slower for $t\simeq45\text{--}54$, where the central values are approximately $0.14$. They approach the lifetime scale but remain above the reference $\Gamma_\tau=0.122$. The displayed uncertainties combine fit covariance and sensitivity to omitting individual scan points; they do not constitute a complete numerical error budget. Later points lie in the region identified as affected by finite-volume effects and are excluded from the interpretation. The comparison demonstrates that the inferred width depends on the observation window as well as on the incoming packet distribution.

\section{Summary and Conclusions}

Real-time scattering is a central target for quantum simulations of field theory. Extending these studies to coupled bosonic and fermionic fields requires suitable quantum encodings and reliable benchmarks for the collision dynamics. With this aim, we have formulated a hybrid qubit--oscillator encoding of lattice Yukawa theory and used MPS simulations to resolve fermion--antifermion collisions in space and time. The joint analysis of scalar excitation, energy exchange and entanglement provides a finite-lattice scattering benchmark. The encoding requires Gaussian scalar evolution, fermionic hopping gates and non-Gaussian conditional displacements as the operations for a photonic implementation.

The collision produces a localised scalar response that is strongest near resonance, suppressed at lower incoming energy and absent in the free theory. Enhanced entanglement accompanies energy transfer from the incoming fermionic sector into the scalar field and its subsequent partial return. Energy scans are compatible with packet-averaged Breit--Wigner profiles. The fitted width of the time-integrated response asymptotes to the analytically predicted width as expected.

These results link tensor-network reference calculations to the development of hybrid quantum simulations of coupled fermionic and bosonic fields. Their scope is a finite-lattice study of collision dynamics. Quantitative extraction of asymptotic scattering parameters 
in future implementation of this approach on genuine quantum hardware would 
require control of initial state preparation.
Methods proposed to address this include 
dressed quasiparticle preparation and detection~\cite{Morgavi2026Quasiparticles}, quantum optimal control~\cite{Araz:2024optimal}, and continuous-variable variational state preparation~\cite{Athanasakos:2026cvadapt}. Combined with the hybrid encoding presented here, these techniques could enable fully dynamical simulations of fermion scattering on photonic quantum computers, opening a route towards studies of real-time, non-perturbative effects in increasingly realistic quantum field theories. \\

\paragraph{Acknowledgements.}
S. A. Abel acknowledges support from STFC under Grant No. ST/X000745/1.

\bibliographystyle{inspire}
\bibliography{refs}{}

\end{document}